\documentclass[aps,onecolumn]{revtex4}

\usepackage[dvips]{epsfig}

\usepackage{subfig}
\usepackage{graphics}
\usepackage{color}
\usepackage{mathrsfs}
\usepackage{amsmath}
\usepackage{amssymb}
\usepackage{float}

\begin{document}

\bibliographystyle{apsrev}

\title
{Importance of fluid inertia for the orientation of 
spheroids settling in turbulent flow} 

\author{Muhammad Zubair Sheikh$^1$, Kristian Gustavsson$^2$, Diego Lopez$^3$, Emmanuel L\'ev\^eque$^3$, Bernhard Mehlig$^2$, Alain Pumir$^1$ and Aurore Naso$^3$}

\affiliation{
$^1$Univ Lyon, ENS de Lyon, Univ Claude Bernard, CNRS, Laboratoire de Physique,\\
F-69342, Lyon, France
\\
$^2$Department of Physics, Gothenburg University, 41296 Gothenburg, Sweden
\\
$^3$Univ Lyon, Ecole Centrale de Lyon, Univ Claude Bernard, CNRS, INSA de Lyon, Laboratoire de M\'ecanique des Fluides et d'Acoustique,
F-69134, \'Ecully, France
}

\begin{abstract}
How non-spherical particles orient as they settle in a flow has important
practical implications in a number of scientific and engineering problems.
In a quiescent fluid, a slowly settling particle orients so that it 
settles with its broad side first. This is an effect of the torque due to 
convective inertia of the fluid set in motion by the settling particle, 
which maximises the drag experienced by the particle. Turbulent flows tend to randomise the particle orientation.
Recently the settling of non-spherical particles in turbulence was analysed neglecting
the effect of convective fluid inertia, but taking into account the effect of the turbulent fluid-velocity gradients on the particle orientation.
These studies reached the opposite conclusion, namely that a rod settles 
preferentially with its tip first, wheras a disk settles with its edge first,
therefore minimizing the drag on the particle.
Here, we consider both effects, the convective inertial torque as well as the torque due to fluctuating velocity gradients, and ask
under which circumstances either one or the other dominate. To this end we estimate the ratio of the magnitudes
of the two torques. Our estimates suggest that  the fluid-inertia torque prevails in high-Reynolds number flows. In this case
non-spherical particles are expected to settle with a maximal drag.
But when the Reynolds number is small then
the torque due to fluid-velocity gradients may dominate, causing the particle to settle with its broad side first.
\end{abstract}

\maketitle

\section{Introduction}
\label{sec:intro}

The settling of non-spherical particles in a flow is of importance in several
scientific and engineering problems~\citep{Bagheri:2016}. 
One example is given by deep cumulus clouds. In such clouds, the
temperature falls well below $0^o C$, inducing the formation of small ice crystals, which
play a very significant role in the 
formation of  precipitation~\citep{PruppKlett,ChenLamb}. 
The orientation 
of such small ice crystals also determines how electromagnetic 
radiation is reflected from clouds~\citep{Yang:15}.
A second example is the sedimentation of
organic and anorganic matter in the turbulent ocean \citep{Kiorboe:01,Ruiz:04}.
The dynamics of motile micro-organisms in turbulence \citep{Dur13,Zha14,Gustav:16,Bor19}, usually slightly heavier than water, 
has important consequences for the population dynamics and the ecology of the ocean system.

More generally,  understanding the angular dynamics of non-spherical particles in turbulent flows
is a challenging fundamental problem~\citep{Voth:2017}. When the particles are heavier than the surrounding fluid,
the problem is significantly complicated.
Even in the simplest case of spherical particles suspended in a turbulent flow, a detailed understanding of their settling 
properties is only beginning to emerge~\citep{Maxey:83,Nielsen:93,Good:14,Bec14,Gus14e}.
Clearly, understanding the angular dynamics is essential to
describe the settling of non-spherical particles.

An exact theoretical description of the problem requires the solution of
the full Navier-Stokes equations, imposing no-slip boundary conditions at the 
surface of the solid~\citep{Naso:10,Homann:10,Candelier19}.
An alternative, much more tractable
approach consists in using simplified descriptions, based on known solutions
of the Navier-Stokes equations.
For example, it is frequently assumed that the particle dynamics can be described by the Stokes approximation, neglecting 
the acceleration of the surrounding fluid set in motion by the particle.
Thus one assumes that the hydrodynamic force on the particle is simply Stokes force, and that the torque is Jeffery's torque \citep{Jeffery:22}. 
The resulting set of equations has been used to study the orientation distribution of non-spherical particles in turbulence \citep{Marchioli10,Gustavsson14,Siewert:14a,Siewert:14b,Gustavsson:17,Jucha:18,Naso:18}. 
This model predicts that the orientation of non-spherical particles settling in turbulence is biased, despite the fact that the turbulence is isotropic. It was found that rods tend to settle tip first, and disks are biased towards settling edge first, i.e. that the orientation of the spheroids preferentially minimises drag.

This model, however, neglects the effect of fluid inertia. How convective fluid inertia affects the orientation of particles settling in a fluid
has been studied mostly for a fluid at rest. Whereas
the corrections to the translational motion, in particular for a sphere, are
very well understood for small particle Reynolds numbers~\citep{Happel83}, the effect of the convective fluid inertia upon the angular dynamics in the same regime has
been less studied.
Analytic expressions were proposed by~\citet{Cox65} for nearly spherical particles, by~\citet{Khayat89} for
slender bodies, and more recently by \citet{Dabade15} for spheroids of arbitrary aspect ratios.

Much less is known when particles settle through fluid in motion.
\citet{Lopez:17} measured the angular distribution of slender rods settling in a two-dimensional (2D) steady vortex flow. In most cases they found that the distributions are strongly peaked so that the rods tend to settle edge first,
as they would in a quiescent fluid. This indicates that the effect of convective fluid inertia dominates the angular dynamics.
\citet{Lopez:17} performed numerical simulations of a model taking into account convective fluid inertia, generally in good agreement with their  experimental results.
\citet{Kramel} analysed the angular dynamics of slender rods
settling in turbulence at different Reynolds numbers, and showed that
rods settle with their broad side first. The observed distribution of orientation is
very narrow, the more so as 
the settling number, $Sv$ (a dimensionless measure of
the settling velocity) is large.
\citet{Klett:95} arrived at the same conclusion, although with
a different prediction for the orientation distribution.
The study of
\citet{Gustavsson:19} provides a consistent description of the orientation
statistics.

The aim of the present work is to determine the conditions under which it may be justified to neglect fluid inertia for the angular dynamics of spheroids settling in turbulence. For this, we use the same model as~\citet{Klett:95} and~\citet{Lopez:17}: we assume
that the torque on the particle is simply the sum of Jeffery's torque and the convective fluid-inertia torque for a particle
moving slowly and steadily in a fluid at rest~\citep{Dabade15}. Our model also includes the first fluid-inertia correction 
due to slip in the centre-of-mass dynamics~\citep{Brenner61}.
We estimate in Section~\ref{sec:rel_torques} the ratio $\mathscr{R}$
between the magnitudes of the convective fluid-inertia torque, $\mathbf{T}_I$, and Jeffery's torque, $\mathbf{T}_{St}$.
The torque induced by fluid inertia therefore dominates when $\mathscr{R}\gg 1$, so that a spheroid settles with its broad side first
(rods with their long edge first, discs broadside down). When $\mathscr{R}\ll 1$, by contrast, we expect a spheroidal particle to settle with its narrow edge
first.

In the case of a laminar flow, with a small Reynolds number $Re_f = U_0 L/\nu$ ($U_0$ and $L$ are respectively the characteristic velocity and lengthscale of the flow, our estimate of 
$\mathscr{R}$ leads to conclusions consistent with those of \cite{Lopez:17},
who studied the settling of rods in the overdamped limit.
In the small flow Reynolds number case, the value of $\mathscr{R}$ can be either
large or small, depending on the ratio between the settling velocity and 
the flow velocity. As a result, the two possible orientation biases (particles
settling with their narrow edge first ($\mathscr{R} \ll 1$), or with their broad
edge first ($\mathscr{R} \gg 1$) can be observed.

In turbulent flows, we derive a different estimate of 
$\mathscr{R}$.
This estimate implies that the convective torque always dominates when
the settling velocity is larger than the typical flow velocity. As the
orientation bias 
obtained when ignoring completely fluid inertia requires the settling velocity 
to be larger than the flow velocity~\citep{Gustavsson:17,Jucha:18}, we 
reach the conclusion that spheroids
settling with their narrow side first, as described in
\citet{Siewert:14a,Siewert:14b,Gustavsson:17,Jucha:18}, cannot be observed
in turbulent flows.

These predictions are in good agreement with results of numerical simulations presented below, in Section~\ref{sec:results}.

\section{Equations of motion }
\label{sec:eqns_motion}

\begin{figure}
\begin{center}
(a)\includegraphics[width=0.36\textwidth]{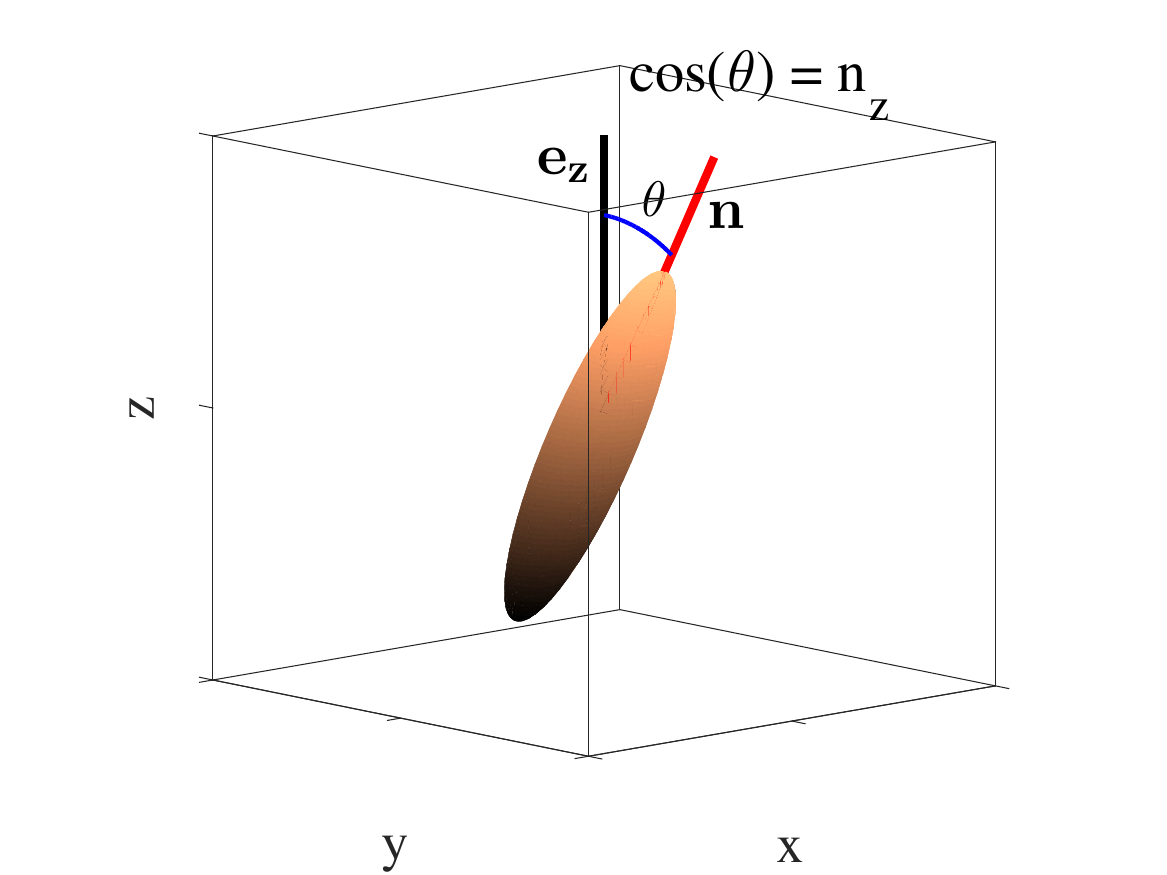}
\hspace{1cm}
(b)\includegraphics[width=0.36\textwidth]{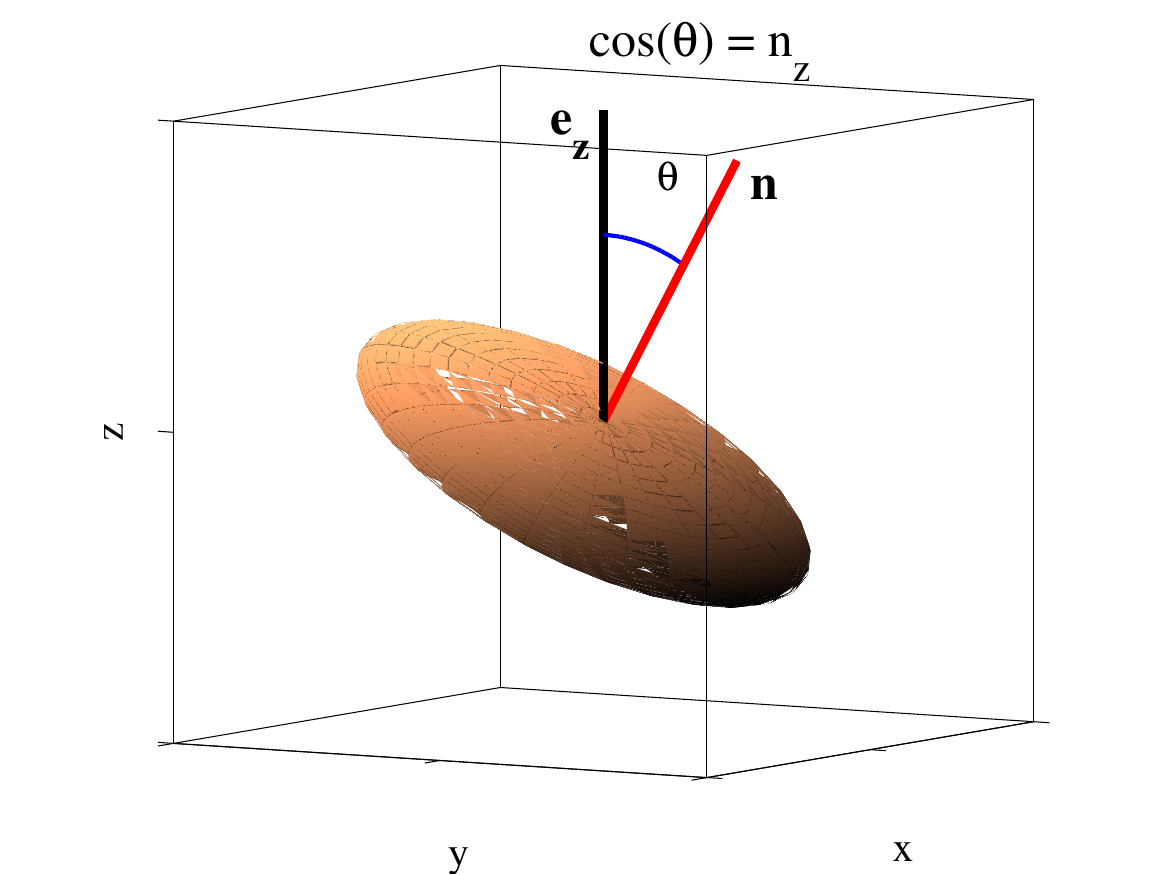}
 \caption{Definition of $n_z$ (projection on the vertical axis of the unit vector aligned with the particle symmetry axis)
 for (a) prolate and (b) oblate particles.}
 \label{fig_def_nz}
\end{center}
\end{figure}

In an incompressible fluid with density 
$\rho_f$ and  kinematic viscosity $\nu=\mu/\rho_f$, we consider solid spheroids of 
density $\rho_p$ and of semi-axes $a$
and $a \beta$. With our convention, oblate (prolate) spheroids correspond to $\beta<1$ 
($\beta > 1$). 
The semi-major axis of the particle, $\tilde{a}$, 
is equal to $a$ ($a\beta$) for oblate (prolate) particles,
and the mass of the particle is 
$m_p = \frac{4}{3} \pi a^3\beta\rho_p$. 
We denote by $\mathbf{n}$ a unit vector aligned with the 
symmetry axis of the particle, $\mathbf{n}$ (Fig. \ref{fig_def_nz}).
The motion of solid particles is entirely characterized by the velocity
of their centers of mass, $\mathbf{v}$, and their angular velocity, 
${\boldsymbol \omega}$.
In the following, we restrict ourselves to particles heavier than the
fluid, $\rho_p \gg \rho_f$.

We restrict ourselves to particles very small compared to the size of the smallest eddies $\eta$, the Kolmogorov length: $\tilde{a}<\eta$.
To describe the dynamics of the settling particle we use the same model as \citet{Klett:95} and~\citet{Lopez:17}.
This model can be rigorously justified only for
a particle  settling in a homogeneous flow at small particle Reynolds number
$Re_p \equiv W_s \tilde{a}/\nu$ \citep{Cox65,Khayat89,Happel83,Dabade15}, and neglecting the effects of shear and unsteadiness 
\citep{Candelier:16}. But the model was shown to qualitatively describe the angular dynamics
of rods settling in a two-dimensional vortex flow \citep{Lopez:17}.
In this model, the equation of motion describing the evolution 
of the particle velocity $\mathbf{v}(t)$ in the laboratory frame reads: 
\begin{equation}
\frac{d\mathbf{v}}{dt} = \mathbf{g}+\frac{6\pi\mu \tilde{a}}{m_p}
\textrm{\bf M}_{St}\cdot(\mathbf{u}-\mathbf{ v})
+\frac{1}{m_p}\frac{9\pi}{8}\rho_f\tilde{a}^2| \mathbf{u}-\mathbf{v}| \textrm{\bf M}_I\cdot(\mathbf{u}-\mathbf{v}),
\label{eq:dvdt}
\end{equation}
where ${\bf g}$ is gravity  and $\mathbf{u}=\mathbf{u}(\mathbf{ x}(t),t)$ represents the undisturbed velocity of 
the fluid at the point where the particle is located, $\mathbf{x}(t)$, at 
time $t$. Aside from gravitational settling, the force acting
on the particle results from the slip velocity, 
$\mathbf{u}_s = \mathbf{v} - \mathbf{u}$.
Together with the equation $ d \mathbf{x}/dt  = \mathbf{v}$, 
Eq.~\eqref{eq:dvdt} determines the position of the particle center of mass. 
The second term on the right-hand side (r.h.s.) of Eq. (\ref{eq:dvdt}) expresses the Stokes drag 
force, appropriate for a spheroidal particle. The expression of this force
involves the anisotropic drag tensor 
\begin{equation}
\textrm{\bf M}_{St}=X^A{\bf n}{\bf n}+Y^A(\mathbb{I}-{\bf n}{\bf n}), \label{eq:def_M_St}
\end{equation}
where ${\bf n}$ is the unit vector
defined above (see Fig. \ref{fig_def_nz}), and $X^A$ and $Y^A$ are dimensionless
coefficients which only depend on the shape of the spheroid through the
parameter $\beta$.
Their expressions can be found in the Appendix.
The third term on the r.h.s. of Eq.\eqref{eq:dvdt} represents the inertial 
contribution to the 
force. The dimensionless tensor $\textrm{\bf M}_I$ takes the form:
\begin{equation}
\textrm{\bf M}_I=[3X^A-(X^A\cos^2\alpha+Y^A\sin^2\alpha)]X^A{\bf n}{\bf n}+[3Y^A-(X^A\cos^2\alpha+Y^A\sin^2\alpha)]Y^A(\mathbb{I}-{\bf n}{\bf n}),
\label{eq:def_M}
\end{equation}
where $\alpha$ is the angle between the slip velocity ${\bf u}_s={\bf v}-{\bf u}$ and ${\bf n}$ ($\alpha=\cos^{-1}({\bf u}_s\cdot{\bf n}/|{\bf u}_s|)$).

The angular equation of motion reads:
\begin{equation}
\frac{d \hat{\boldsymbol{\omega}}}{dt}=\begin{pmatrix} \hat{T}_{xx}/\hat{I}_{xx} \\ \hat{T}_{yy}/\hat{I}_{yy} \\ \hat{T}_{zz}/\hat{I}_{zz} \end{pmatrix}+\begin{pmatrix} \hat{\omega}_y\hat{\omega}_z\frac{\hat{I}_{yy}-\hat{I}_{zz}}{\hat{I}_{xx}} \\ \hat{\omega}_z\hat{\omega}_x\frac{\hat{I}_{zz}-\hat{I}_{xx}}{\hat{I}_{yy}} \\ \hat{\omega}_x\hat{\omega}_y\frac{\hat{I}_{xx}-\hat{I}_{yy}}{\hat{I}_{zz}} \end{pmatrix}. \label{eq:rot_2}
\end{equation}
We expressed here the equations
of motion in the particle frame $( \hat{x}, \hat{y}, \hat{z})$ with axes $\hat{\bf e}_x$, $\hat{\bf e}_y$, $\hat{\bf e}_z$
attached to the particle. Hereafter, all the quantities expressed in 
this frame carry a hat.
The relation between a tensor expressed in the reference frame
of the particle, $\hat{\mathbf{K}}$, and in the reference frame of the 
laboratory, $\mathbf{K}$ is given by: 
$\hat{\mathbf{K}} = \mathbf{R^{rot}} \, \mathbf{K} \, \mathbf{R^{rot}}^{-1}$, where 
$\mathbf{R^{rot}}$ is the rotation matrix between the two frames.
We note that the equation of evolution of $\mathbf{R^{rot}}$ is simply
$d R_{ij}^{rot}/dt = \varepsilon_{ikl} \Omega_{k} R_{lj}^{rot}$, where $\varepsilon_{ijk}$ is the Levi-Civita symbol. In our simulations, rotations were represented in terms of quaternions \citep{Jucha:18,Naso:18,Almondo18}.
The tensor $\hat{\mathbf{I}}$ is the moment-of-inertia tensor, which is 
diagonal in the reference frame of the particle $(\hat{x}, \hat{y}, \hat{z})$:
$\hat{\mathbf{I} } = (4 \pi \rho_p a^5 \beta/15)  \, {\rm diag}( 1 + \beta^2, 1 + \beta^2, 2)$. 
The second term on the r.h.s. of Eq.~\eqref{eq:rot_2} 
represents the torque induced by the non-Galilean nature of the 
particle
frame~\citep{LL_mech}. The first term on the r.h.s. of Eq.~\eqref{eq:rot_2}
results from the hydrodynamic torque $\hat{\mathbf{T}}$ acting on the
particle. 

Following \citet{Lopez:17}, we express 
$\hat{\mathbf{T}}$ as the sum of Jeffery's torque~\citep{Jeffery:22}, $\hat{\mathbf{T}}_{St}$, and of the 
contribution due to fluid inertia~\citep{Cox65,Khayat89,Dabade15}, $\hat{\mathbf{T}}_I$ for a particle
moving steadily in a homogeneous flow:
$\hat{\mathbf{T}} = \hat{\mathbf{T}}_{St} + \hat{\mathbf{T}}_I$.  
Jeffery's torque is given by~\citep{Jeffery:22}:
\begin{equation}
\hat{\bf T}_{St}=\frac{16}{3}\pi\mu a^3\beta\begin{pmatrix} \frac{1+\beta^2}{\alpha_0+\beta^2\gamma_0} & 0 & 0 \\ 0 & \frac{1+\beta^2}{\alpha_0+\beta^2\gamma_0} & 0 \\ 0 & 0 & \frac{1}{\alpha_0} \end{pmatrix} \cdot \begin{pmatrix} \frac{1-\beta^2}{1+\beta^2}\hat{S}_{zy}+(\hat{\Omega}_{zy}-\hat{\omega}_x) \\ \frac{\beta^2-1}{1+\beta^2}\hat{S}_{xz}+(\hat{\Omega}_{xz}-\hat{\omega}_y) \\ \hat{\Omega}_{yx}-\hat{\omega}_z \end{pmatrix} ,
\label{eq:Jeff_torq}
\end{equation}
where $\hat{S}_{ij}$ and $\hat{\Omega}_i$ are the components of the strain 
and vorticity, expressed in the reference frame of the particle. The 
quantities $\alpha_0$ and $\gamma_0$ are dimensionless, and depend 
only on $\beta$, see Appendix.
The torque induced by convective fluid inertia reads, derived in
\citep{Dabade15}, reads (to order $Re_p$):
\begin{equation}
\hat{\bf T}_{I}=\mathbf{ R^{rot}} {\bf T}_{I}, ~~ {\rm with} ~~
{\bf T}_{I}=-\rho_f |{\bf u}-{\bf v}|^2\tilde{a}^3F_\beta \left( \frac{{\bf u}_s}{|{\bf u}_s|}\cdot{\bf n} \right) \left( \frac{{\bf u}_s}{|{\bf u}_s|}\times{\bf n} \right) ,
\label{eq:I_torq}
\end{equation}
where $F_\beta$ is a dimensionless shape factory which depends only on $\beta$. Its expression
is given in the Appendix.
We have neglected here possible fluid-inertia corrections due to shear.
This is justified 
provided that the Oseen length, ${\ell}_O \approx \nu/W_s$, is much smaller than 
the Saffman length, ${\ell}_S \approx (\nu/s)^{1/2}$, where $s$ is the order
of magnitude of the velocity gradients in the flow~\citep{Candelier19}. 
We must therefore require that
${\ell}_S/{\ell}_O \gg 1$,  an assumption we reconsider in the following
Section.

\section{Relative importance of Jeffery's torque and fluid-inertia torque}
\label{sec:rel_torques}

To estimate the relative importance of the inertial effects on the dynamics of
the particles orientation, we calculate the ratio 
$\mathscr{R}\equiv |\hat{\bf T}_I|/|\hat{\bf T}_{St}|$.
The particles orientational dynamics is therefore expected to 
be correctly described by the Jeffery torque when $\mathscr{R}\ll 1$.
We begin by looking at the case of very slim, prolate spheroids, with 
$\beta\gg 1$. In this limit, one obtains:
\begin{eqnarray}
\hat{T}_{I,x}/\hat{T}_{St,x} \sim \hat{T}_{I,y}/\hat{T}_{St,y}\sim\frac{5u_s^2}{8\nu s \log\beta},\\
\hat{T}_{I,z}/\hat{T}_{St,z}\sim\frac{5u_s^2\beta^2}{16\nu s(\log\beta)^2}, \label{eq:TI_TsT_fibres}
\end{eqnarray}
where we recall that $\mathbf{u}_s$ is the slip velocity,
and where $s$ is the inverse of a characteristic time scale representing the angular slip velocity $|\boldsymbol{\Omega}-\boldsymbol{\omega}|$ or the local strain $|\bf{S}|$.
To derive the estimate for the ratio between $\hat{T}_I$ and $\hat{T}_{St}$ in Eq.~\eqref{eq:TI_TsT_fibres}, we used the fact that $F_\beta$ 
is no larger than $1$ for fibers with $\beta \gg 1$, 
see Fig.\ref{fig:parameters}.
The terms responsible for the change in orientation are the
$\hat{x}$ and $\hat{y}$ components in Eq.~\eqref{eq:TI_TsT_fibres}. For the present purpose, we therefore
consider only these components.
As we are interested only in orders of magnitude, we will approximate 
$5/8$ by $1$ in the following, and the $\log(\beta)$ factor, which
in practice is close to $1$ if $5 \le \beta \le 100$, will be omitted.

In the opposite limit of very thin disks, $\beta \ll 1$, one obtains:
\begin{equation}
\hat{T}_I/\hat{T}_{St} \sim \frac{3}{32} F_0 \frac{u_s^2}{\nu s}, \label{eq:TI_TSt_disques}
\end{equation}
where $u_s$ and $s$ are defined as before, and 
$F_0 = \lim_{\beta \rightarrow 0} F_\beta =38/9-17216/(945\pi^2)$ \citep{Dabade15}, so 
$3F_0/32 \sim 1$. 
The factor $F_\beta$ has once again been omitted since its absolute value is always smaller than $2.5$, as shown in Fig. \ref{fig:parameters}.
Equation (\ref{eq:TI_TSt_disques}) for disks therefore only differs from 
Eq. (\ref{eq:TI_TsT_fibres}) for rods by the $1/\log(\beta)$ factor, which
in practice does not vary much, as mentioned in the previous paragraph.

To summarise, for very flat disks ($\beta\ll 1$) and for thin rods ($5 \le \beta \le 100$, not too thin so that the value of $\log\beta$ remains moderate),
the ratio $\hat{T}_I/\hat{T}_{St}$ is of order $\mathscr{R}$, defined by
\begin{equation}
|\mathbf{T}_{St}|/|\mathbf{T}_I|\sim \mathscr{R} ,~~{\rm with} ~~~~ \mathscr{R} \equiv \frac{| \mathbf{u} - \mathbf{v} |^2 }{\nu | {\boldsymbol \Omega} - {\boldsymbol \omega} | }  \, ,
\label{eq:def_R}
\end{equation}
where $\mathbf{v}$ is the particle
velocity, ${\boldsymbol \omega}$ its angular velocity, $\mathbf{u}$ is the fluid velocity,
${\boldsymbol \Omega}$ the vorticity and $\nu$ is the viscosity of the fluid.
The differences $| \mathbf{u} - \mathbf{v} |$ and 
$ | {\boldsymbol \Omega} - {\boldsymbol \omega}| $ are the 
translational and rotational slip velocities. The convective fluid inertia torque dominates when $\mathscr{R}\gg 1$, in this case
a spheroid settles with maximal drag. When $\mathscr{R}\ll 1$, by contrast, we expect a spheroid to settle with minimal drag.
To provide an estimate for $\mathscr{R}$, we approximate the slip velocity as $u_s \sim W_s$, where $W_s \approx g \tau_p$ is the settling velocity 
of the particle. Here $\tau_p$ is an estimate of its response time, defined as $a^2\log\beta(\rho_p/\rho_f)/(3\nu)$ for prolate bodies and $\pi a^2\beta(\rho_p/\rho_f)/(8\nu)$ for oblate ones (strictly speaking, the particle linear response time depends on its orientation, as stated in Eq.~(\ref{eq:dvdt})).
To estimate the inverse characteristic time scale $s$, we need
to distinguish the laminar and turbulent cases. 

\subsection{Turbulent flows}

Turbulent flows  generate large gradients. Standard estimates
show that the vorticity root-mean square is of order 
$|\langle {\boldsymbol \Omega}^2 \rangle^{1/2}| \sim (U_0/L) Re_f^{1/2}$. 
Substituting this expression, together with $| \mathbf{u} - \mathbf{v} | 
\sim W_s$ in Eq.~\eqref{eq:def_R} leads to:
\begin{equation}
\mathscr{R} \sim \Big( \frac{W_s}{U_0} \Big)^2 Re_f^{1/2}\,.
\label{eq:R_turb}
\end{equation}
Using the Kolmogorov velocity scale~\citep{Frisch95}, $u_K \sim U_0 Re_f^{-1/4}$,
we can rewrite this estimate 
as $\mathscr{R} \sim ({W_s}/{u_K})^2$, consistent with the formulation in \cite{Gustavsson:19} where the width
of the orientation distribution in the limit
$\mathscr{R}\gg 1$ is evaluated.
In any case, Eq.~(\ref{eq:R_turb}) shows that in the high $Re_f$ regime, the fluid-inertia torque can be neglected (\textit{i.e.}, $\mathscr{R}$ can be small) only if $W_s/U_0$ is small. In this limit, the orientation distribution is approximately uniform, as shown in \citep{Siewert:14a,Siewert:14b,Gustavsson:17,Jucha:18}.
This means that the orientation bias analysed in these works (preferential minimal drag) 
cannot be actually observed at large $Re_f$.

The estimate for the velocity gradients, $s \sim 1/\tau_K$,
leads to the ratio between the Saffman and the Oseen lengths~\citep{Candelier19}
${\ell}_S/{\ell}_O\sim \mathscr{R}^{1/2}\sim W_s/u_K$.
This demonstrates that neglecting 
fluid-inertia effects due to shear, as done here, is consistent when $\mathscr{R} \gg 1$.\\

\subsection{Laminar flows}
\label{subsec:lam_flow}

The analysis developed in the previous subsection applies also to a laminar flow, such as the cellular flow considered by \citet{Lopez:17}. In the case of a laminar flow, at small $Re_f$, the velocity gradient can be simply estimated as $s\sim U_0/L$, so that:
\begin{equation}
\mathscr{R}\sim \Big( \frac{W_s}{U_0} \Big)^2  Re_f .
\label{eq:R_lam}
\end{equation}
We remark that the relative magnitude of the fluid-inertia contributions  to the angular dynamics [Appendix B of~\citep{Lopez:17}]
amounts to
$Re_p w_0/\lambda_p$, where $Re_p = \beta a W_s/\nu$, $w_0 = W_s/U_0$ and $\lambda_p = \beta a/L$. As a 
result, our expression ~(\ref{eq:R_lam}) coincides with theirs.
Equation~(\ref{eq:R_lam}) suggests that the distribution of orientation predicted
in~\citep{Siewert:14a,Siewert:14b,Gustavsson:17,Jucha:18,Naso:18} --
spheroids settling preferentially with a minimal drag --
could be observable at small $Re_f$.

We notice that estimating $\mathscr{R}$ in the limit $Re_f\rightarrow 0$ 
requires some care. 
To fully analyze the problem, it is convenient to introduce, in
addition to $U_0$ and $W_s$,
the velocity $V_\nu = \nu/L$, where $L$ is the characteristic length scale 
of the problem and $\nu$ the viscosity. The two dimensionless parameters 
$Re_f$ and $\mathscr{R}$ can simply be expressed as: $Re_f = (U_0/V_\nu)$ and 
$\mathscr{R}=(W_s/U_0) \times (W_s/V_\nu)$. The limit $Re_f \rightarrow 0$ may be 
reached in two very different ways, either by letting $U_0 \rightarrow 0$ or
$V_\nu \rightarrow \infty$. In the former case, 
$U_0 \rightarrow 0$ at fixed $V_\nu$, the parameter $\mathscr{R}$ is very large,
so the fluid inertia dominates. In contrast, letting 
$V_\nu \rightarrow \infty$ at fixed $U_0$ leads to $\mathscr{R}\rightarrow 0$, so the
Jeffery torque prevails. A careful estimate of the ratios
between $W_s$, $U_0$ and $V_\nu$ is therefore necessary to conclude about the 
importance of fluid inertia in the $Re_f \rightarrow 0$ limit.

To summarize, the main result of this section is 
our systematic analysis of
the ratio between the torque induced
by fluid convection and the Jeffery torque, and quantified
by $\mathscr{R}$, introduced by Eq.~\eqref{eq:def_R}. With our definition, 
the torque induced
by convective forces dominates when $\mathscr{R}\gg 1$. A more refined relation
between $\mathscr{R}$, the Reynolds number $Re_f$, and the ratio
between the settling velocity, $W_s$ and the flow 
velocity, $U_0$, depends on whether the flow is turbulent, 
Eq.~\eqref{eq:R_turb}, or laminar, Eq.~\eqref{eq:R_lam}.\\

\begin{figure}
\begin{center}
(a)\includegraphics[width=0.27\textwidth]{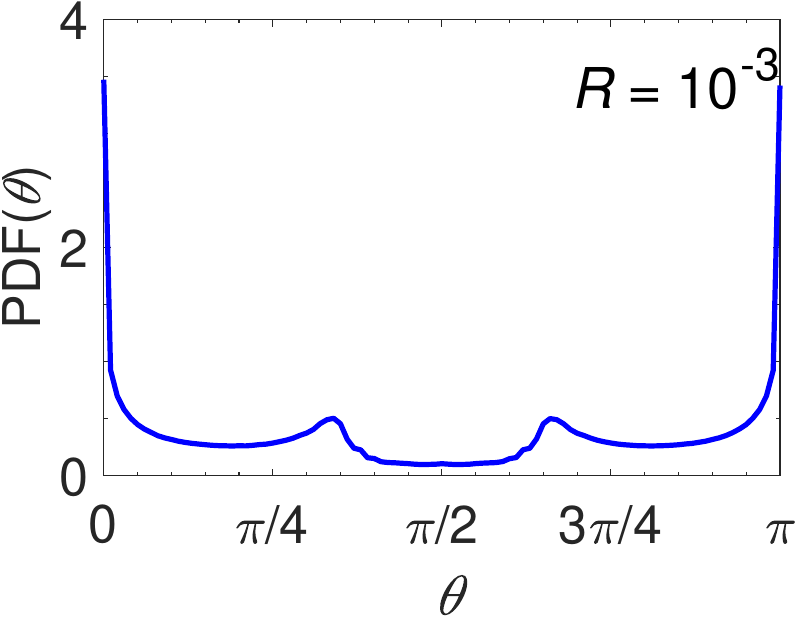}
\hspace{0.2cm}
(b)\includegraphics[width=0.27\textwidth]{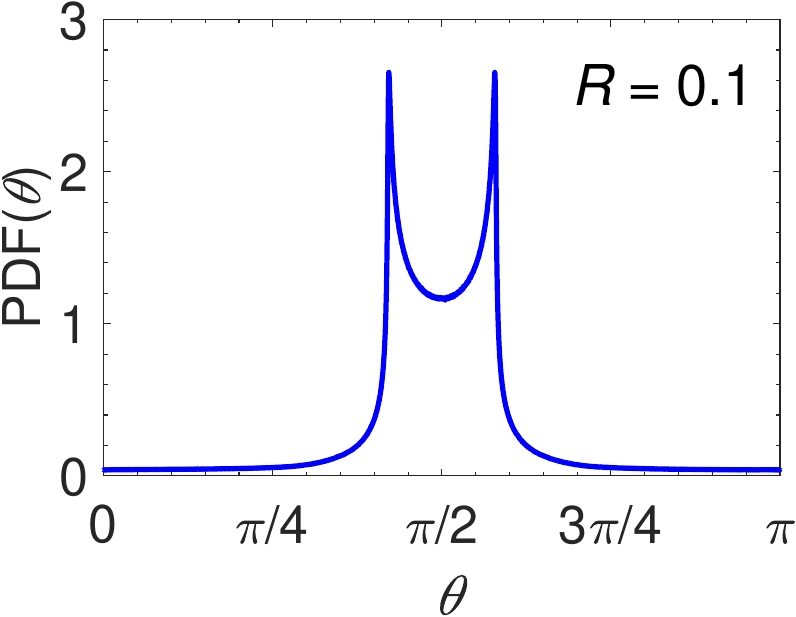}
\hspace{0.2cm}
(c)\includegraphics[width=0.27\textwidth]{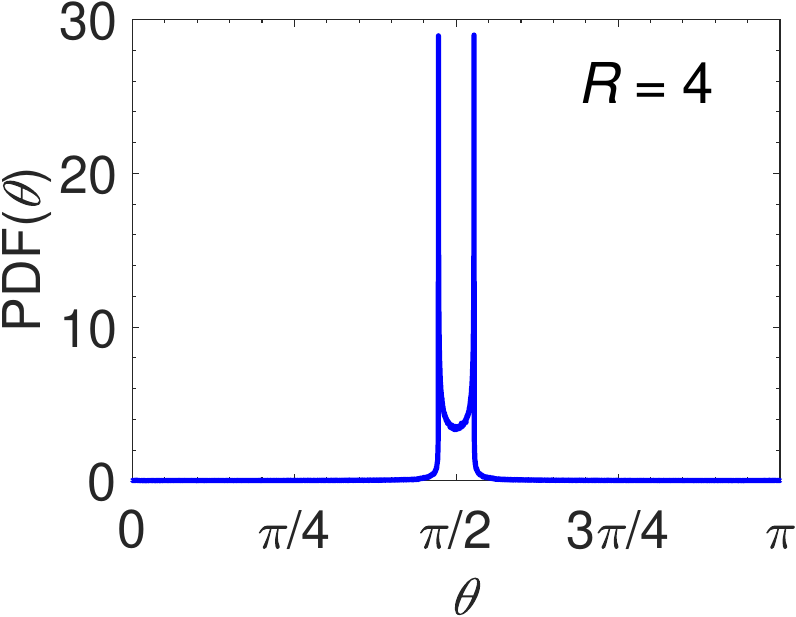}

    \caption{Distribution of $\theta$ (angle between the axis of symmetry of the particle and gravity) for prolate bodies of aspect ratio $\beta=20$ in the 2D vortex flow: (a) $\mathscr{R}=10^{-3}$ ($W_s/U_0 \approx 0.283$, $Re_f=1/80$), (b) $\mathscr{R}=0.1$ ($W_s/U_0 \approx 2.88$, $Re_f=1/80$) and (c) $\mathscr{R}=4.0$ ($W_s/U_0 \approx 6.32$, $Re_f=0.1$). A uniform distribution of $\theta$ would correspond to a random orientation of the particle.
    }
	\label{fig:2DTG_PDF_nz}
\end{center}
\end{figure}

\section{Results of numerical simulations}
\label{sec:results}
In this section, we test numerically the results of the analysis of the
previous paragraph. To this end,
we study the dynamics of spheroids in different flow models that describe  low-$Re_f$ flows, as well as high-$Re_f$ (turbulent) flows with many spatial scales.

\subsection{Two-dimensional vortex flow}

We performed numerical simulations of prolate spheroids ($\beta=20$) settling in a
  simple two-dimensional (2D) cellular flow 
mimicking that of~\cite{Lopez:17}. 
The (marginal) probability distribution function of $\theta$, the angle between the axis of symmetry of the particle and gravity, is shown in Fig.~\ref{fig:2DTG_PDF_nz} for different values of $\mathscr{R}$.
In this 2D flow, a random orientation of the particle would result in a uniform distribution of $\theta$.
The distributions shown in Fig.~\ref{fig:2DTG_PDF_nz} are symmetric with respect to the value $\pi/2$, as expected due the symmetry of the objects.
At low values of $\mathscr{R}$ (Fig.~\ref{fig:2DTG_PDF_nz}a), the orientation  
distribution 
exhibits a peak around $\theta \approx 0$ and $\theta \approx \pi$.
Here $\theta$ is the angle between  $\mathbf{n}$ and the direction
of gravity (Fig.~\ref{fig_def_nz}). So this corresponds
 to rods  settling with their tips first. As the value of $\mathscr{R}$
increases (Fig.~\ref{fig:2DTG_PDF_nz}b, c), however, the peaks migrate towards $\theta \approx \pi/2$.
This corresponds to rods settling with their broad sides first.

\subsection{Three-dimensional turbulence}

\begin{figure}
    \centering
(a)\includegraphics[width=0.41\textwidth]{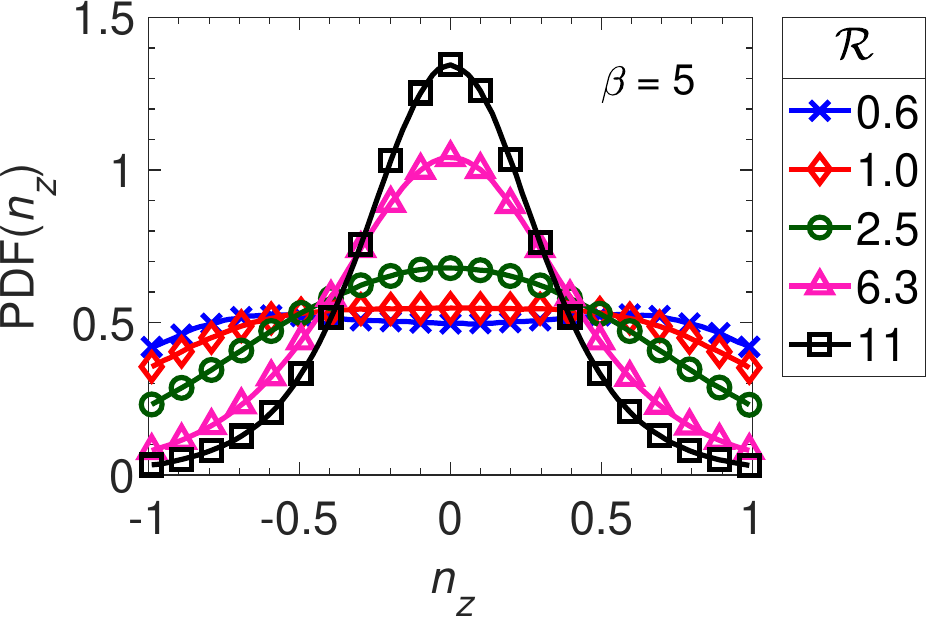}
\hspace{1cm}
(b)\includegraphics[width=0.41\textwidth]{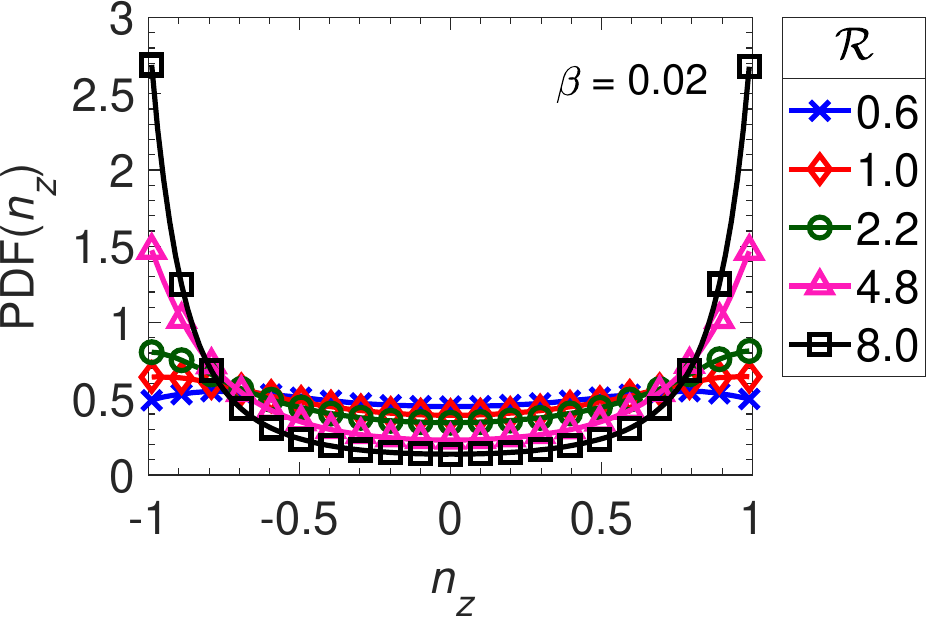}
    \caption{Distribution of $n_z$ for (a) prolate ($\beta=5$) and (b) oblate ($\beta=0.02$) particles in the 3D turbulent flow (DNS), for $St=0.1$ and several values of $\mathscr{R}$. In this 3D flow, a random orientation of the particles would be reflected by a uniform distribution of $n_z$.
    }
\label{fig:allR}
\end{figure}

We investigate turbulent flows
 a cubic box with periodic boundary conditions in all directions. Our
pseudo-spectral (de-alised) code, see e.g. \citep{Jucha:18}, is fully 
de-aliased. We used a
grid resolution $N^3=128^3$, 
corresponding to a Reynolds number based on the Taylor microscale $R_\lambda=40$. An external force acts on low-wavenumber modes to ensure a constant injection rate of energy, $\varepsilon=100\; cm^2.s^{-3}$, and reach stationary homogeneous and isotropic turbulence. The kinematic viscosity $\nu=0.15\; cm^2.s^{-1}$ has been adjusted to get a satisfying grid resolution $k_{max}\eta\sim 3.4$, where $k_{max}$ is the highest wave number and $\eta=(\nu^3/\varepsilon)^{1/4}$ is the Kolmogorov scale.

\begin{figure}
    \begin{center}
(a)\includegraphics[width=0.41\textwidth]{R3_p_legend.pdf}
\hspace{1cm}
(b)\includegraphics[width=0.41\textwidth]{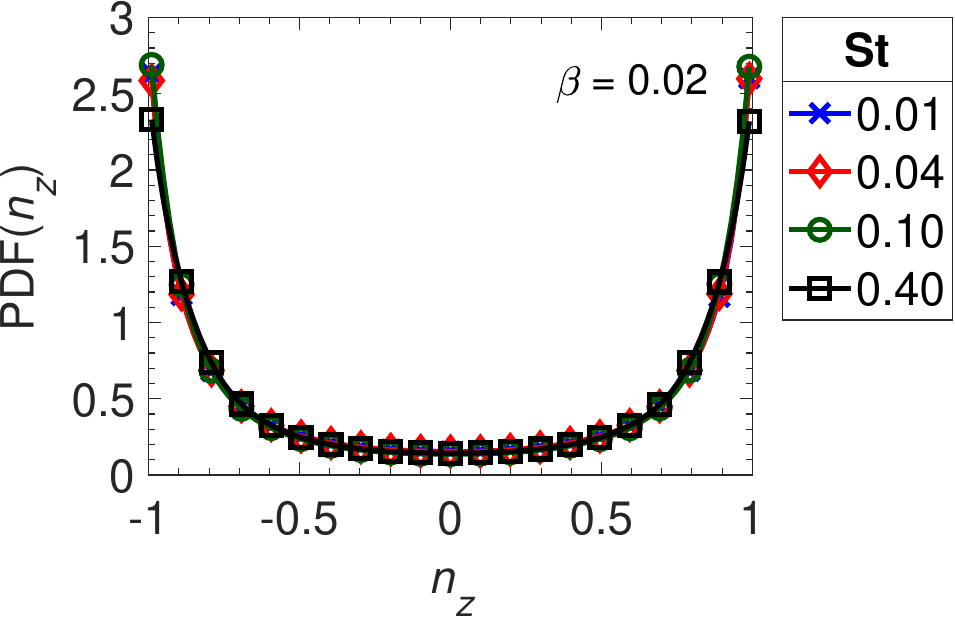}
    \caption{Distribution of $n_z$ for  $\mathscr{R}\approx10$ and several values of the Stokes number, for (a) prolate ($\beta=5$) and (b) oblate ($\beta=0.02$) particles in the 3D turbulent flow.}
\label{fig:R10}
    \end{center}
\end{figure}

Figures~\ref{fig:allR} and \ref{fig:R10} display the results of our simulations.
Shown is the distribution of $n_z$, the projection on the vertical axis of the unit vector aligned with
the particle symmetry axis (see Fig. \ref{fig_def_nz}). In 3D flows, such a distribution would be uniform if the spheroids were randomly oriented.
The distribution of $n_z$ is first shown in Fig.~\ref{fig:allR} for different values of $\mathscr{R}$,
for prolate spheroids ($\beta = 5$), panel a, and for oblate spheroids 
($\beta = 0.02$), panel b. These results were obtained for a constant value of the 
Stokes number $St = \tau_p/\tau_K$, where $\tau_K$ is the Kolmogorov time scale.

As expected, Fig. \ref{fig:allR} shows that for $\mathscr{R}\lesssim 1$ the distribution
of angles is almost uniform.
When
the parameter $\mathscr{R}$ becomes
larger than $1$,
the  fluid-inertia torque dominates the angular dynamics.
For prolate particles (Fig.~\ref{fig:allR}(a))
the distribution peaks at $n_z \approx 0$,
where $\mathbf{n}\perp\mathbf{g}$. This
corresponds to particles settling with their long side first. 
For oblate spheroids, on the other hand (Fig.~\ref{fig:allR}(b)),
the distributions peak
at $n_z = \pm 1$. This corresponds to disks settling with their broad sides first.
For increasing values of $\mathscr{R}$ (not shown here), the distributions of $n_z$ become more and more peaked around $n_z=0$ for prolate spheroids, and $n_z=\pm 1$ for oblate particles.

Our simulations results
fully support the prediction that
when the settling parameter 
$w_0=W_s/U_0$ is of order unity or larger, then $\mathscr{R} > 1$ in a turbulent flow
with large $Re_f$. In this case the fluid-inertia torque dominates the angular dynamics,
leading to a distribution of orientation maximising the drag, at odds with the predictions of models
neglecting this torque \citep{Siewert:14a,Siewert:14b,Gustavsson:17,Jucha:18,Naso:18}.
A theory for the orientation distribution for $\mathscr{R} > 1$, when the 
fluid-inertia torque dominates, has been derived by~\citet{Gustavsson:19} 
(see also \citep{Kramel}).

Figure \ref{fig:R10} shows that at a fixed 
value of $\mathscr{R}$, the Stokes number has only a 
limited effect on the orientation statistics, for the range of parameters 
considered here ($St \lesssim 1$),
consistent with the findings of \citet{Gustavsson:19} (our numerical data exhibit the same behavior for any value of $\mathscr{R}$).\\

Tables~\ref{table:allR} and \ref{table:R10}
list the parameters used in the simulations of Fig.~\ref{fig:allR} and \ref{fig:R10}, respectively.
We defined the
settling numbers $Sv_L=g\tau_p/U_0$, $Sv_\eta=g\tau_p/u_K$, and velocity ratio $w_0 = W_s/U_0$.
The particle Reynolds number, $Re_p$, is defined 
as $W_s \tilde{a}/\nu$. The parameter $S=|{\bf u}-{\bf v}|/W_s$ is the ratio between the slip velocity and the settling one. Its value is always close to one, thereby confirming the fact that $|{\bf u}-{\bf v}|$ is, in the present setup, always well approximated by $W_s$.

\hspace{-1in}

\begin{table}
\subfloat{
\centering
\begin{tabular}{l|ccccccc}
        &$\mathcal{R}$&$w_o$&$\mathrm{Re}_p$&$Sv_{L}$&$Sv_{\eta}$&$S$\\
        \hline
        1&0.6&0.21&0.029&0.17&0.31&1.08\\
        2&1.0&0.36&0.047&0.32&0.57&1.13\\
        3&2.5&0.57&0.074&0.50&0.87&1.15\\
        4&6.3&0.88&0.117&0.87&1.55&1.14\\
        5&11&1.13&0.147&1.12&2.01&1.12\\
        \end{tabular}
} \qquad
\subfloat{
\begin{tabular}{l|ccccccc}
        &$\mathcal{R}$&$w_o$&$\mathrm{Re}_p$&$Sv_{L}$&$Sv_{\eta}$&$S$\\
        \hline
        1&0.6&0.20&0.051&0.20&0.35&1.04\\
        2&1.0&0.37&0.084&0.36&0.64&1.15\\
        3&2.2&0.54&0.123&0.56&1.00&1.14\\
        4&4.8&0.81&0.181&0.87&1.54&1.20\\
        5&8.0&1.03&0.222&1.13&2.02&1.20\\
        \end{tabular}
}
\caption{
Parameters corresponding to Fig.~\ref{fig:allR}: (left) prolate particles, see Fig.~\ref{fig:allR}a; (right) oblate particles, see Fig.~\ref{fig:allR}b. $St=0.1$.}
\label{table:allR}
\end{table}

\begin{table}
\subfloat{
\centering
\begin{tabular}{l|cccccccc}
        &$\mathcal{R}$&St&$w_o$&$\mathrm{Re}_p$&$Sv_{L}$&$Sv_{\eta}$&$S$\\
        \hline
        1&10&0.01&1.08&0.047&1.12&2.01&1.07\\
        2&10&0.04&1.06&0.091&1.09&1.95&1.10\\
        3&11&0.10&1.13&0.147&1.12&2.01&1.12\\
        4&12&0.40&1.28&0.332&1.25&2.23&1.12\\

        \end{tabular}
} \qquad
\subfloat{
\begin{tabular}{l|cccccccc}
        &$\mathcal{R}$&St&$w_o$&$\mathrm{Re}_p$&$Sv_{L}$&$Sv_{\eta}$&$S$\\
        \hline
        1&7.0&0.01&0.96&0.065&1.06&1.89&1.15\\
        2&7.5&0.04&0.975&0.129&1.07&1.90&1.18\\
        3&8.0&0.10&1.03&0.222&1.13&2.02&1.20\\
        4&8.0&0.40&1.12&0.439&1.16&2.06&1.14\\
        \end{tabular}
}
\caption{
Parameters corresponding to Fig.~\ref{fig:R10}: (left) prolate particles, see Fig.~\ref{fig:R10}a; (right) oblate particles, see Fig.~\ref{fig:R10}b.}
\label{table:R10}
\end{table}

\begin{figure}
    \centering
(a)\includegraphics[width=0.36\textwidth]{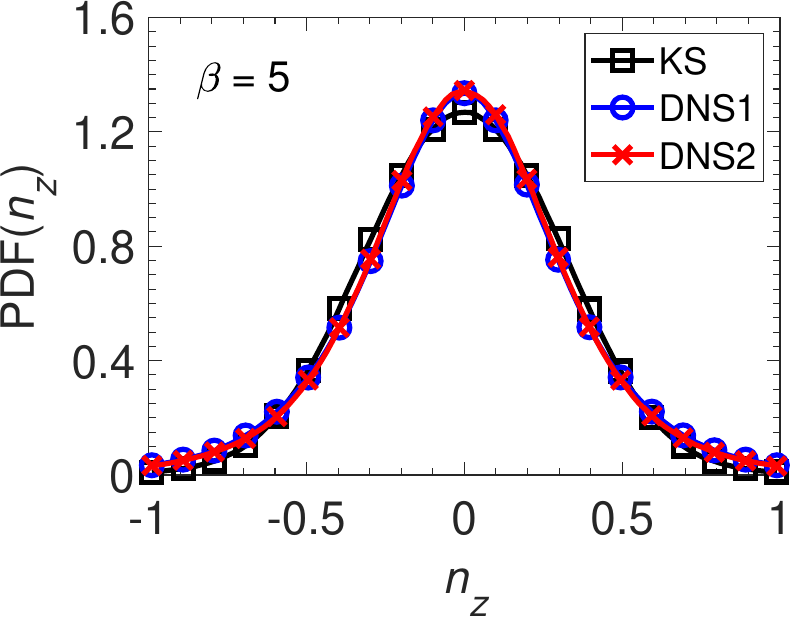}
\hspace{1cm}
(b)\includegraphics[width=0.36\textwidth]{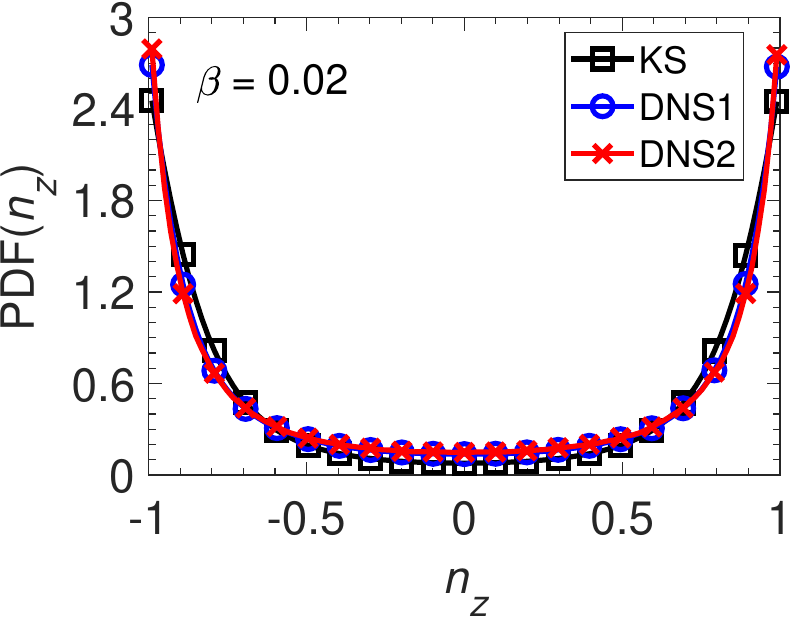}
    \caption{Comparison of distribution of $n_z$ calculated by Kinematic Simulation (KS) and two runs of Direct Numerical Simulation (DNS), $\mathcal{R}\approx10$ and St=0.1: (a) prolate ($\beta=5$) and (b) oblate ($\beta=0.02$) particles. DNS1 and KS: $R_\lambda=40$; DNS2: $R_\lambda=80$.
    }
\label{fig:KS_DNS}
\end{figure}

The previous results were obtained for a moderate value of $R_\lambda=40$. As shown in Fig.~\ref{fig:KS_DNS}, similar results are obtained for $R_\lambda=80$. Finally, we stress the fact that, for these very dense particles, similar results can also be obtained by using much simplified models of
turbulent flows.
In fact, we compared directly the prediction of the Kinematic
Simulation (KS) model of turbulence with the results of DNS.
Briefly, the KS flow~\citep{Fung:92} is obtained by
representing
the large range of scales in the flow by a superposition of a few Fourier modes, with a distribution of 
wavenumber amplitude in $k$-space. Details about our implementation can
be found in~\citep{Ducasse:10}.
As shown in 
Fig. \ref{fig:KS_DNS}, we found a very good agreement between the DNS
and KS results,
as it was the case in \citep{Gustavsson:17}. 
This can be understood by noticing that the velocity gradient 
tensor acting on a settling particle
decorrelates faster than on a tracer particle, and 
as a result, the particle response is not sensitive to the fine correlations
of the velocity gradient.\\

\section{Discussion} \label{sec:concl}
The analysis presented here, along with the numerical results obtained using
very simplified flows, shows that the effect of convective fluid inertia 
on the orientation of particles settling in a flow 
becomes important when the parameter 
$\mathscr{R}$, defined by Eq.~\eqref{eq:def_R} is $\gtrsim 1$. This 
condition implies
$(W_s/U_0)^2 \, Re_f^{\alpha} \gtrsim 1$, where $\alpha = 1$ in the
laminar regime ($Re_f < 1$), and $\alpha = 1/2$ at large Reynolds 
numbers, when the flow is turbulent. 

For a turbulent flow this condition implies that the effect of convective fluid inertia can only be neglected when 
$W_s/U_0 \ll 1$. In this regime, the orientation distribution is 
essentially uniform. When the ratio $W_s/U_0$ becomes of order unity, our study 
suggests that a particle tends to settle preferentially broad side down
(with a maximal drag orientation).
This conclusion has been previously reached by considering the restoring torque
induced by convective inertia~\citep{Klett:95}, and has also
been observed in laboratory experiments \citep{Kramel}. 
The expression for $\mathscr{R}$ given by Eq.~\eqref{eq:def_R}
was derived in the limits of infinitely thin ($\beta\gg 1$) or infinitely flat 
($\beta\ll 1$) particles, but our numerical results show that this parameter is also relevant for 
moderate values of $\beta$.

The present analysis was performed by keeping, in addition to the
Jeffery torque, only the leading-$Re_p$
contribution of the convective fluid-inertia torque for a
particle settling slowly and steadily in a fluid at rest.
This model has
been validated, to some extent, in a recent experimental work~\citep{Lopez:17}.
This opens the possibility to understand quantitatively
the orientation distribution of small non-spherical particles settling 
in turbulent flows~\citep{Gustavsson:19},  a question of fundamental importance
in cloud physics~\citep{PruppKlett}.

\acknowledgments{
We acknowledge discussions with E. Guazzelli and J. Magnaudet. This work was supported by 
the IDEXLYON project (Contract ANR-16-IDEX-0005) under
University of Lyon auspices.
We also aknowledge support from Vetenskapsr\aa{}det 
[grant number 2013-3992], Formas [grant number 2014-585], and from the 
Knut and Alice Wallenberg Foundation 
(grant \lq Bottlenecks for particle growth in turbulent aerosols\rq{}  2014.0048).
The simulations were performed using the resources provided by PSMN (ENS Lyon).
}

\section*{Appendix A: Expression of the coefficients $X^A$, $Y^A$, $\alpha_0$, $\gamma_0$ and $F_\beta$}
\label{appendix:A}

In the following, we introduce the ellipticity $e$, defined for oblate 
bodies ($\beta < 1$) by $ e  =  \sqrt{1 - \beta^2 } $
and for prolate ones ($\beta > 1$) by
$ e =  \sqrt{1 - (1/\beta)^2 }$.

\bigskip
\noindent\textit{Expression of the force}
We begin by giving the expressions of the coefficients $X^A$ and $Y^A$
that define the resistance tensor $\textrm{\bf M}_{St}$, determined  
in~\cite{Kim91} and introduced in
Eq.~\eqref{eq:def_M_St},\eqref{eq:def_M}.
For prolate spheroids ($\beta>1$):
\begin{equation}
X^A=\frac{8e^3}{3 \left( -2e+(1+e^2)\ln \left( \frac{1+e}{1-e} \right) \right) },
\quad
Y^A=\frac{16e^3}{3 \left( 2e+(-1+3e^2)\ln \left( \frac{1+e}{1-e} \right) \right) }.
\end{equation}
For oblate spheroids ($\beta<1$):
\begin{equation}
X^A=\frac{4e^3}{3 \left( (2e^2-1)\cot^{-1} \left( \frac{\sqrt{1-e^2}}{e} \right) +e\sqrt{1-e^2} \right) },
\quad
Y^A=\frac{8e^3}{3 \left( (2e^2+1)\cot^{-1} \left( \frac{\sqrt{1-e^2}}{e} \right) -e\sqrt{1-e^2} \right) }.
\end{equation}

\begin{figure}
    \begin{center}
\includegraphics[width=0.61\textwidth]{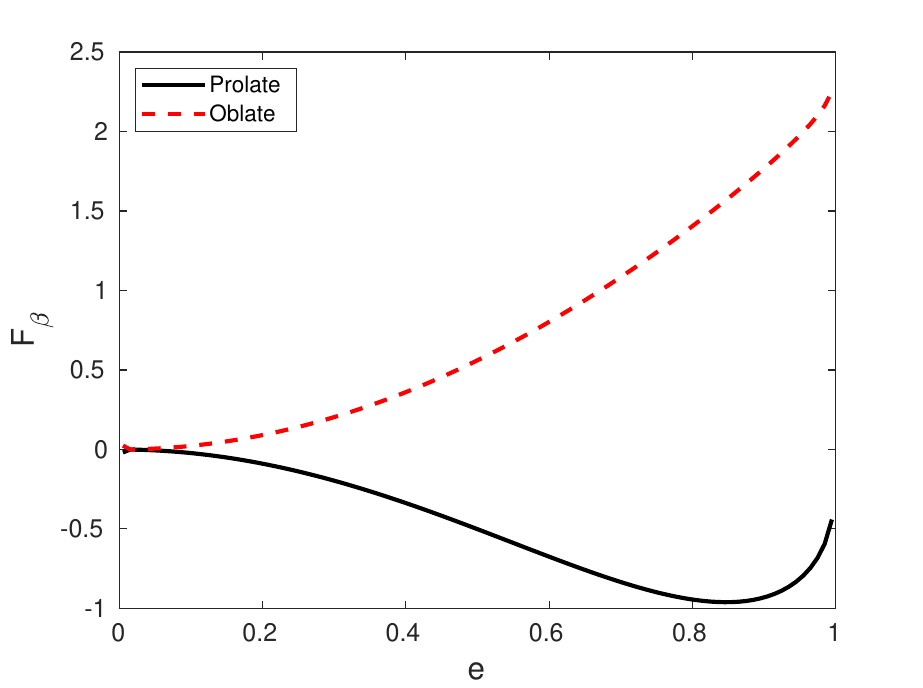}
    \caption{Parameter $F_\beta$ plotted as a function of the particle ellipticity, for prolate and oblate spheroids.}
\label{fig:parameters}
    \end{center}
\end{figure}

\bigskip
\noindent\textit{Expression of the torque}
The tensor necessary to calculate the Jeffery torque, see 
Eq.~\eqref{eq:Jeff_torq}, requires the two parameters $\alpha_0$ and 
$\gamma_0$ \citep{Jeffery:22}. Additionally, the expression for the torque due to inertial
effects involves a factor $F_\beta$ (see Eq.~\eqref{eq:I_torq}), derived by \citet{Dabade15} (Eq.~(4.1) and (4.2) therein).
The expressions of these parameters are given below, and the $\beta$-dependence of $F_\beta$ is more specifically shown in Fig. \ref{fig:parameters}. \\
For prolate spheroids ($\beta>1$):
\begin{equation}
\alpha_0=\frac{\beta^2}{\beta^2-1}-\frac{\textrm{arccosh}(\beta)\beta}{(\beta^2-1)^{3/2}},
\quad
\gamma_0=\frac{-2}{\beta^2-1}+\frac{2\textrm{arccosh}(\beta)\beta}{(\beta^2-1)^{3/2}},
\end{equation}

\begin{equation}
\begin{split}
F_\beta  =&\frac{-\pi e^2(420e+2240e^3+4249e^5-2152e^7)}{315((e^2+1)\tanh^{-1}e-e)^2((1-3e^2)\tanh^{-1}e-e)}\\
&+\frac{\pi e^2(420+3360e^2+1890e^4-1470e^6)\tanh^{-1}e}{315((e^2+1)\tanh^{-1}e-e)^2((1-3e^2)\tanh^{-1}e-e)}\\
&-\frac{\pi e^2(1260e-1995e^3+2730e^5-1995e^7)(\tanh^{-1}e)^2}{315((e^2+1)\tanh^{-1}e-e)^2((1-3e^2)\tanh^{-1}e-e)}.
\end{split}
\end{equation}
For oblate spheroids ($\beta<1$):
\begin{equation}
\alpha_0=\frac{\beta^2}{\beta^2-1}+\frac{\arccos(\beta)\beta}{(1-\beta^2)^{3/2}},
\quad
\gamma_0=\frac{2}{1-\beta^2}-\frac{2\arccos(\beta)\beta}{(1-\beta^2)^{3/2}},
\end{equation}

\begin{equation}
\begin{split}
F_\beta =&\frac{\pi e^3\sqrt{1-e^2}(-420+3500e^2-9989e^4+4757e^6)}{315\sqrt{1-e^2}(-e\sqrt{1-e^2}+(1+2e^2)\sin^{-1}e)(e\sqrt{1-e^2}+(2e^2-1)\sin^{-1}e)^2}\\
&+\frac{210\pi e^2(2-24e^2+69e^4-67e^6+20e^8)\sin^{-1}e}{315\sqrt{1-e^2}(-e\sqrt{1-e^2}+(1+2e^2)\sin^{-1}e)(e\sqrt{1-e^2}+(2e^2-1)\sin^{-1}e)^2}\\
&+\frac{105\pi e^3(12-17e^2+24e^4)(\sin^{-1}e)^2}{315(-e\sqrt{1-e^2}+(1+2e^2)\sin^{-1}e)(e\sqrt{1-e^2}+(2e^2-1)\sin^{-1}e)^2}.
\end{split}
\end{equation}

\end{document}